\def\figsize{\hsize}

\documentclass[draft,grl]{AGUTeX}

 \usepackage[pdftex]{graphicx}
\authorrunninghead{YIZHAQ ET AL.}

\titlerunninghead{SCALE FREE DISTRIBUTION OF SINKHOLES }

\authoraddr{Corresponding author: H. Yizhaq,
Department of Solar Energy and Environmental Physics, Ben-Gurion
University, Midreshet Ben-Gurion 84990, Israel. (yiyeh@bgu.ac.il)}

\begin{document}

%% ------------------------------------------------------------------------ %%
%
%  TITLE
%
%% ------------------------------------------------------------------------ %%

\title{Scale-free distribution of Dead Sea sinkholes--observations and modeling}
%
% e.g., \title{Terrestrial ring current:
% Origin, formation, and decay $\alpha\beta\Gamma\Delta$}
%

%% ------------------------------------------------------------------------ %%
%
%  AUTHORS AND AFFILIATIONS
%
%% ------------------------------------------------------------------------ %%

%Use \author{\altaffilmark{}} and \altaffiltext{}

% \altaffilmark will produce footnote;
% matching \altaffiltext will appear at bottom of page.

 \authors{H. Yizhaq \altaffilmark{1,2}, C. I. Shalom \altaffilmark{2},
 E. Raz \altaffilmark{2}, and Y. Ashkenazy \altaffilmark{1}}

\altaffiltext{1}{Blaustein Institutes for Desert Research, Ben-Gurion University of the Negev, Sede Boqer Campus 84990, Israel.}

\altaffiltext{2}{The Dead Sea and Arava Science Center, Tamar Regional Council, Israel.}

%\altaffiltext{3}{The Dead Sea and Arava Science Center, Tamar Regional Council, Israel.}

%\altaffiltext{4}{Blaustein Institutes for Desert Research, Ben-Gurion University of the Negev, Sede Boqer Campus 84990, Israel.}

%\altaffiltext{5}{Dipartimento di Idraulica, Trasporti ed
%Infrastrutture Civili, Politecnico di Torino, Turin, Italy.}

%% ------------------------------------------------------------------------ %%
%
%  ABSTRACT
%
%% ------------------------------------------------------------------------ %%

% >> Do NOT include any \begin...\end commands within
% >> the body of the abstract.

\begin{abstract}
  There are currently more than 5500 sinkholes along the Dead Sea in Israel. These were formed due to the dissolution of subsurface salt layers as a result of the replacement of hypersaline groundwater by fresh brackish groundwater. This process has been associated with a sharp decline in the Dead Sea water level, currently more than one meter per year, resulting in a lower water table that has allowed the intrusion of fresher brackish water. We studied the distribution of the sinkhole sizes and found that it is scale-free with a power-law exponent close to 2. We constructed a stochastic cellular automata model to understand the observed scale-free behavior and the growth of the sinkhole area in time. The model consists of a lower salt layer and an upper soil layer in which cavities that develop in the lower layer lead to collapses in the upper layer. The model reproduces the observed power-law distribution without involving the threshold behavior commonly associated with criticality.
\end{abstract}

%% ------------------------------------------------------------------------ %%
%
%  BEGIN ARTICLE
%
%% ------------------------------------------------------------------------ %%

% The body of the article must start with a \begin{article} command
%
% \end{article} must follow the references section, before the figures
%  and tables.

\begin{article}

%% ------------------------------------------------------------------------ %%
%
%  TEXT
%
%% ------------------------------------------------------------------------ %%

\section{Introduction}\label{sec:sec1}
\begin{figure}
  \begin{center}
  \includegraphics*[width=\figsize]{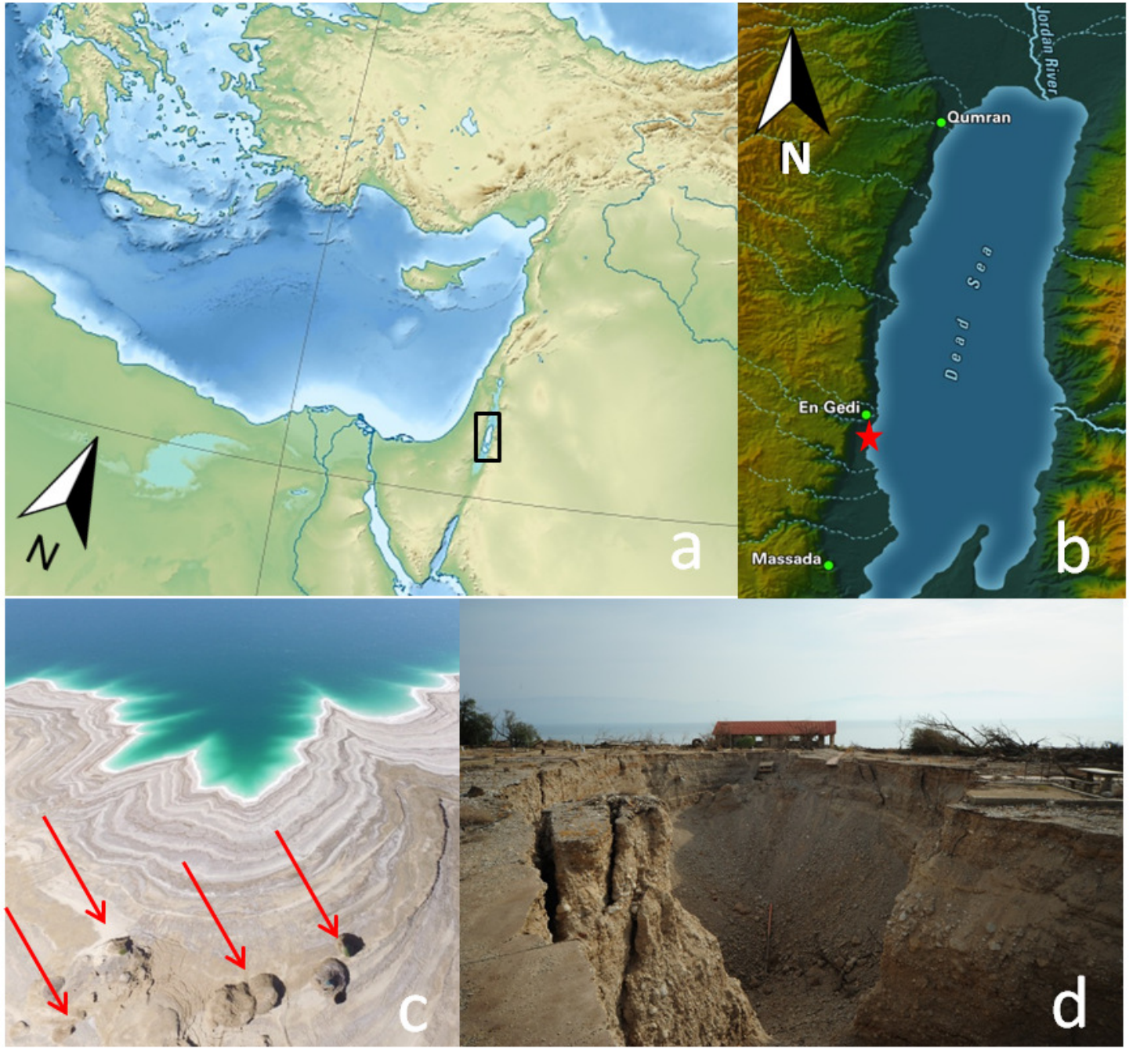}
  \end{center}
  \caption{
    (a) The eastern Mediterranean Sea: the location of the Dead Sea is indicated by the black rectangle. (b) The Dead Sea and its surrounding area. (c) A cluster of sinkholes [indicated by the red star in (b)] located south of Kibbutz Ein-Gedi. (d) A large sinkhole in the alluvial fan near the tourist site of Kibbutz Ein-Gedi. The sinkholes in this area have caused serious damage to the tourist infrastructure; the tourist camp in this region was eventually abandoned.
    \label{fig:fig1}}
\end{figure}

Sinkhole formation along the Dead Sea shorelines (Fig.~\ref{fig:fig1} and Figs. S1-S4) was first observed in 1980, but since then, their formation rate has accelerated due to the ongoing sharp decline in the Dead Sea's water level, currently more than one meter per year~\cite[][]{Arkin-Gilat-2000:dead, Filin-Baruch-Avni-Marco-2011:sinkhole}; the Dead Sea level is presently about 430~m below mean sea level. This sharp decline is mainly attributed to human activities such as the interception of the freshwater supply from the Jordan River and the maintenance of large evaporation ponds by the Dead Sea mineral industries in Jordan and Israel~\cite[][]{Arkin-Gilat-2000:dead}. The sinkhole formation process first began in the southern part of the Dead Sea coast and spread northward along the western (Israeli) coast. The steeper eastern (Jordanian) coast has been less affected, with most of its sinkholes concentrated in the flat-lying region close to the Lisan Peninsula in the southern part of the Dead Sea~\cite[][]{Frumkin-Ezersky-Al-Zoubi-Akkawi-Abueladas-2011:dead, Kottmeier-Agnon-Al-Halbouni-2016:new}. Since 2013, about 400 new sinkholes have appeared each year causing damage to infrastructure, palm plantations, and tourist facilities. As they endanger the stability of the present infrastructure, they pose a severe threat to future regional development along the Dead Sea.

The formation of sinkholes along the Dead Sea is attributed to the collapse of the soil layer (gravel and clay) into underground cavities that were formed by the dissolution of subsurface salt layers due to the replacement of hypersaline groundwater by undersaturated brackish groundwater~\cite[][]{Yechieli-2000:fresh}. The depth of the top of the salt layer ranges between 20 and 50 m (depending on the distance from the Dead Sea); in some locations, the thickness of the salt layer exceeds 20 m and its age is $\sim 10000$ years~\cite[][]{Yechieli-1993:effects}. As the subsurface solutional voids enlarge, a successive roof collapse occurs, depending on the mechanical ability of the overlying layers to withstand the increasing shear stress with respect to the void proportions~\cite[][]{Frumkin-Raz-2001:Collapse}. Thus, the Dead Sea sinkholes have been characterized as "collapse sinkholes"~\cite[][]{Gutierrez-Parise-DeWaele-Jourde-2014:structure}. There are about 60 clusters of sinkholes across a 60-km-long strip along the western shore of the Dead Sea ~\cite[][]{Shalev-Lyakhovsky-2012:viscoelastic}. Recent studies~\cite[][]{Abelson-Yechieli-Crouvi-et-al-2006:collapse, Shalev-Lyakhovsky-Yechieli-2006:salt} showed that some of the sinkholes are located along lineaments, indicating  young, permeable faults that serve as fluid conduits for groundwater to flow through the salt layer. The sinkholes along the Dead Sea developed on two main sedimentary environments~\cite[][]{Shalev-Lyakhovsky-2012:viscoelastic}: mud-flats, consisting of fine-grained sediments, and alluvial fans, mainly consisting of gravel alternating with fine-grained sediments. This difference is expressed in the sinkholes' morphology; the alluvial fan sinkholes are deep (up to 30 m), while the mud-flat sinkholes start from shallow, wide structures that slowly grow in size as their depths also increase. It was suggested~\cite[][]{Shalev-Lyakhovsky-Yechieli-2006:salt} that the areas and the rate of sinkhole formation depend on several properties, such as the surface area, the permeability of the salt and clay layers beneath the freshwater aquifer, the permeability-porosity relation, the dispersivity, and the thickness of the layers.

\section{Scale-free distribution of the Dead Sea sinkhole area}\label{sec:sec2}
Despite the many studies on the Dead Sea sinkholes~\cite[][]{Atzori-Antonioli-Salvi-Baer-2015:InSAR, Kottmeier-Agnon-Al-Halbouni-2016:new}, there are almost no studies regarding the size distribution of the sinkholes and their areal coverage's time development.

Power-law distributions or scale-free patterns occur in an extraordinarily diverse range of phenomena in physical systems, such as the size distributions of both earthquakes~\cite[][]{Steacy-McCloskey-1999:heterogenity} and moon craters~\cite[][]{Newman-2005:power,Clauset-Shalizi-Newman-2009:power}, as well as in various other types of phenomena. Scale-free patterns have also been found in ecological systems, including mussel beds, forest gaps, forest fires and vegetation patchiness in water-limited systems ~\cite[][]{Scanlon-Caylor-Levin-Rodriguez-Iturbe-2007:positive, Kefi-Rietkerk-Alados-Pueyo-et-el:role, Meron-2015:nonlinear}.
Mathematically, the power law of a physical quantity or probability distribution $y(x)$ can be described by
\begin{equation}\label{eq:power_law}
  y(x) \propto x^{ - \beta } \quad{\rm  for}\quad x > {x_{\min }} \,,
\end{equation}
where $\beta$ is the scaling exponent, and ${x_{\min }}$ is a minimum value of $x$ above which the power-law distribution holds~\cite{Stumpf-Porter-2012:critical}. A power-law distribution should exhibit an approximately linear relationship on a log-log plot over at least two orders of magnitude in both the $x$ and $y$ directions~\cite[][]{Stumpf-Porter-2012:critical}.

\begin{figure}
  \begin{center}
  \includegraphics*[width=\figsize]{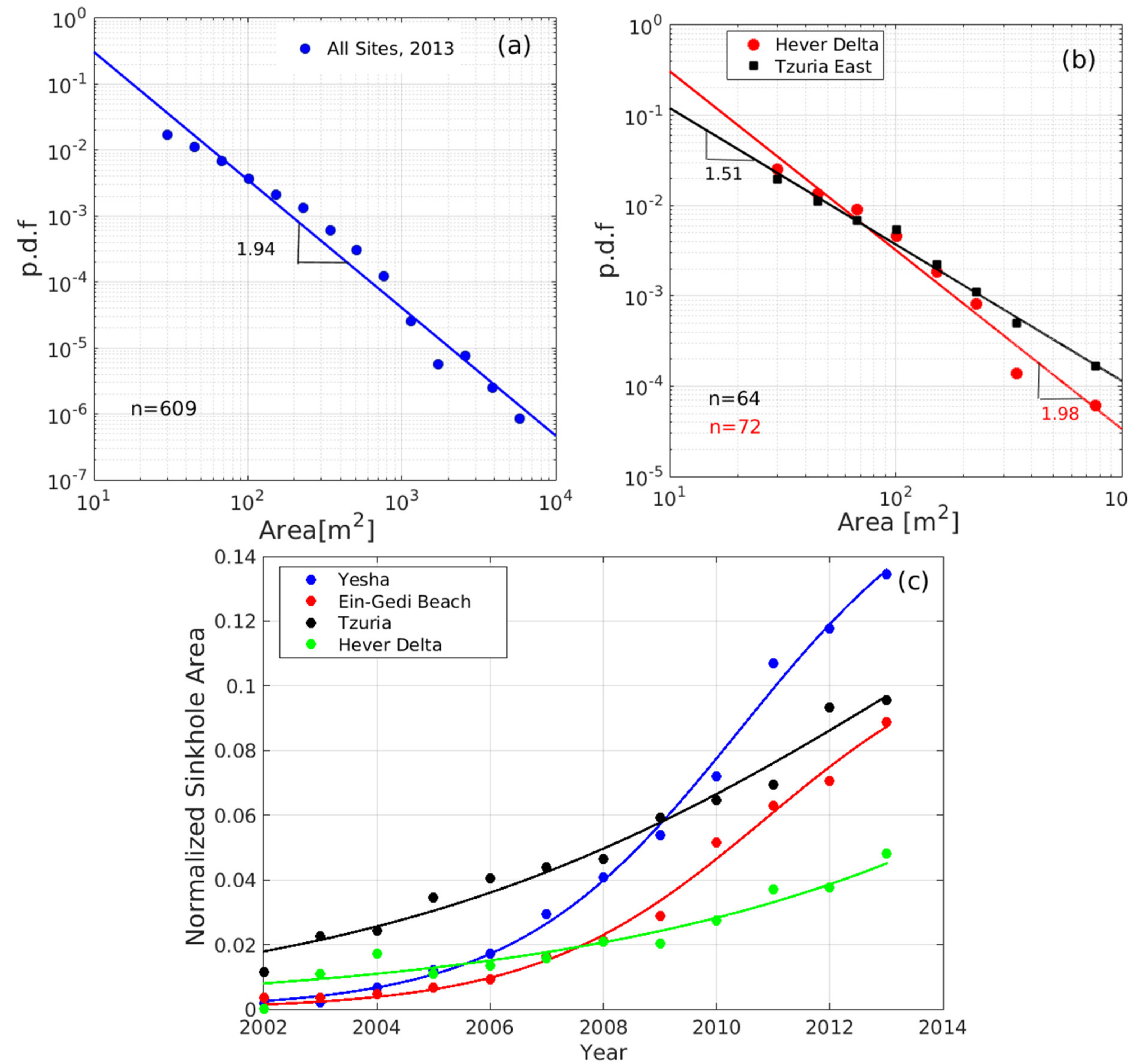}
  \end{center}
  \caption{
    (a) Probability density functions of sinkhole areas for 2013 (for the 609 sinkholes larger than 20 $\rm{m^2}$). The scaling exponent is $\beta=1.94\pm0.09$. (b) Same as (a) for two sinkhole clusters (south of Kibbutz Ein-Gedi, Fig. \ref{fig:fig1}b), Hever Delta and Tzuria East; the scaling exponents are $\beta=1.98\pm0.14 $ and $\beta=1.51\pm0.07$, respectively. (c) Sinkhole area as a function of time (2002-2013) for four sites (opaque circles) and a fit following ~Eq.~(\ref{eq:cover}) (solid lines). The area of a site is defined as the area of a polygon that is tangential to all the circles with a radius of 30 m, measured from each one of the sinkholes inside the site (see Fig. S1)}.
    \label{fig:fig2}
\end{figure}

Since 2002, the evolution and morphometry of 12 sinkhole sites have been monitored either monthly, seasonally or yearly (Figs. S5-S6), depending on the site activity and the level of risk; in 2013, these 12 sites included more than 1300 sinkholes. For each of the 12 sites, the morphological changes (including size and depth) of all the sinkholes were measured in situ. All the sinkhole clusters used in the current study are located south of Kibbutz Ein-Gedi on the western shoreline of the Dead Sea (see Fig.~\ref{fig:fig1}). The probability density function (pdf) of the size distribution of all the sites' sinkholes (with an area larger than 20 $\rm{m^2}$) follows a power law with an exponent $\beta=1.94$ (Fig. 2a), spanning over almost three orders of magnitude of the sinkhole area. Fig. 2b depicts the pdf of two specific sites, Hever Delta (alluvial fans, Fig. S1) and Tzuria East (dried mud, Fig. S2), for which their corresponding scaling exponents are $\beta=1.98$ and $\beta=1.51$, respectively. The difference in the scaling exponent between different sites is attributed to the soil characteristics, as well as to other effects including floods (which affect sinkholes located near wadis), and to the precipitation frequency and magnitude that can increase the amount of fresh water in the salt layer that comes from the sinkhole itself. In addition, the depth of the salt layer, which is a function of the distance of the site from the Dead Sea and the local elevation, may affect sinkhole development. Thus, different sites may be at different stages of their development, associated with different scaling exponents. Generally, it seems that the older the site, the smaller its scaling exponent; this observation is consistent with the model presented below. For example, the Tzuria site is more developed than the Hever Delta site since its normalized area is larger (Fig.~\ref{fig:fig2}c); therefore, its scaling exponent is smaller than the Hever Delta site (Fig.~\ref{fig:fig2}b).

The scaling exponents can also be calculated by the method suggested by~\cite{Clauset-Shalizi-Newman-2009:power}:
 \begin{equation}\label{eq:beta}
\beta  = 1 + n\left[ {\sum\limits_{i = 1}^n {\ln \left( {\frac{{x_i }}{{x_{\min } }}} \right)} } \right]^{ - 1} \,,
\end{equation}
where $n$ is the number of observations, and $x_i$ is the sinkhole area (in $\rm{m^2}$). The standard error of the scaling exponent, which is derived from the width of the log-likelihood function around the maximum~\cite[][]{Clauset-Shalizi-Newman-2009:power}, is given by:
 \begin{equation}\label{eq:sigma}
\sigma  = \frac{{\beta  - 1}}{{\sqrt n }} + {\rm O}(1/n)\,.
\end{equation}

Using these two equations, we obtained the following scaling exponents: $1.69\pm0.07$, $1.78\pm0.22$ and $1.97\pm0.23$, for all-sites, Tzuria East, and Hever Delta, respectively (see also Fig. S8 for the time dependence of $\beta$). Generally speaking, the exponents obtained by the two methods (slope on a log-log plot and ~\cite[][]{Clauset-Shalizi-Newman-2009:power}) are similar. It is important to note that, usually, due to limited observations, it is difficult to prove real scale-free phenomena. In addition, frequently, scale-free phenomena hold only for a certain range of scales, as different processes can be dominant on different scale ranges, resulting in crossovers in the ``scaling'' curves.

The normalized sinkhole area as a function of time, $s(t)$, is shown in Fig. 2c for four sites and can be approximated by a hyperbolic tangent function,
 \begin{equation}\label{eq:cover}
s(t) = a_1 (\tanh (a_2 (t - a_3 )) + 1)\,,
\end{equation}
where $a_1$, $a_2$ and $a_3$ are fitting parameters. The idea behind this fitting function is that in the initial stages, the sinkhole area is very small and then it increases due to internal feedbacks. Then, after a sufficiently long time, the sinkhole area becomes so large that its growth slows down, eventually stopping when it covers the entire site area ($s=1$) since there are no more empty sites left in which new sinkholes can develop. This hyperbolic tangent is not the only possible choice, and a one parameter Holling Type III function can be another option as shown in (Figs. S9 and S10).

\section{The model}\label{sec:sec3}
To understand the scale-free distribution of the sinkhole area, we constructed a simple phenomenological cellular automata model that is based on the basic known physical processes that control the formation and development of sinkholes. Cellular automata models, in which complex system dynamics are represented by simple rules of interaction, have been commonly used to study power-law clustering in natural systems ~\cite[][]{Scanlon-Caylor-Levin-Rodriguez-Iturbe-2007:positive, Kefi-Rietkerk-Alados-Pueyo-et-el:role}. The model does not aim to describe the 3d nature of cavity formations in the salt layer nor the effect of geologic faults on the spatial distribution of sinkholes~\cite[][]{Shalev-Lyakhovsky-2012:viscoelastic}. The main goal of the model is to study whether the basic known mechanisms of sinkholes can produce the observed scale-free behavior and the evolution of the sinkhole area.

The model consists of two 2d layers denoted as lower and upper. The lower layer represents the salt layer where dissolving cavities evolve; by ``cavity,'' we refer to a collection of connected individual empty grid points (or ``holes''). The top layer represents the soil stratum that collapses when the cavity in the lower layer beneath becomes large enough. The coupling between the two layers is unidirectional, i.e., cavities in the lower layer affect the upper layer but not vice versa. The dynamic of the lower layer is governed by the following rules. In each time step, there is a certain probability, $p$, for a site to become a hole, i.e., the salt in this site is dissolved. The evolution of the cavity is modeled by a diffusion-like process in which the probability of a cell (grid point) to become a ``hole'' depends on the number of ``holes'' among its four nearest neighbors and on the cavity size itself. This probability is updated in each time step by $dp$ defined by:
\begin{equation}\label{eq:dp}
dp =  dp_{\max } (1 - \exp ( - \gamma A))\,,
\end{equation}
where $dp_{\max } = (1 - p)/4$ such that the probability of a cell (grid point) to become a hole cannot exceed 1. [For the cases in which the four neighbors of a grid point are holes and the cavity area $A$ is large ($A\rightarrow\infty$), $p+dp_{\max }=p+4(1-p)/4=1$.] $\gamma$ is a parameter that captures the effectivity of salt dissolution. The larger $\gamma$ is, the larger the probability that a certain cell will become a hole. $dp$ also depends on the cavity area since the dissolution of salt increases the porosity and the permeability of the salt layer and initiates the development of dissolution channels~\cite[][]{Weisbrod-Alon-Mordish-Konen-Yechieli-2012:dynamic}, thereby increasing both the rate of solute transport and the rate of the chemical reaction~\cite[][]{Shalev-Lyakhovsky-Yechieli-2006:salt}. This positive feedback underlies the ``reactive infiltration instability''~\cite[][]{Ortoleva-Chadam-Merino-Sen-1987:geochemical, Aharonov-Spiegelman-Kelemen-1997:three} that accelerates the dissolution process. Thus, in the formulated model, the larger the cavity, the faster its growth rate. In each time step, this size condition is checked over all the cavities in the lower layer and updated accordingly.

The dynamics in the upper layer is dictated by the state of the lower layer. The cells in the upper layer that are above a cavity will collapse with a probability $p_u$ defined as:
  \begin{equation}\label{eq:pA}
    p_u = 1 - \exp ( - \alpha r)\,,
\end{equation}
where $r=A/(L+W)$; $L$ and $W$ are the width and length of the bottom cavity's dimensions. The parameter $\alpha$ is related to the properties of the soil layer, and a larger value of $\alpha$ implies a higher probability to collapse. The collapse process also depends on the shape and area of the cavity. The motivation for this form of $r$ comes from the fact that smaller and more elongated cavities provide greater support for the overlying soil layer \cite[][]{Atzori-Antonioli-Salvi-Baer-2015:InSAR}. The probability to collapse becomes larger when the cavity below is larger. The initial state of the lower layer is random where each site in the lower layer is assigned a probability $p$ to become a cavity. The initial state of the upper layer is uniform with no sinkholes at all, and it evolves due to the lower layer state according to the rules defined in Eq.~\ref{eq:pA}. The initial probability distribution is set for $(n+2)\times(n+2)$ grid points, whereas the evolution of the cavities (lower layer) and the sinkholes (upper layer) are calculated only for a slightly smaller $n\times n$ grid; the two extra points in each direction are used to satisfy the periodic boundary conditions.

In summary, the model evolves as follows:
\begin{enumerate}
\item Randomly assigning holes in the lower layer with a probability $p$.
\item Finding cavities (i.e., clusters of continuous holes) in the lower layer and then their areas, and calculating $dp$ for each cavity according to Eq.~(\ref{eq:dp}).
\item Assigning probability $p$ to all the cells in the lower layer and then, for each cell in the existing cavities, adding $dp$ for each of its four neighbors.
\item Finding the dimensions of each cavity in the lower layer and updating the collapse probability in the upper layer according to Eq.~(\ref{eq:pA}).
\item Updating the lower layer: a cell will become a hole (empty) with the probability calculated in step (iii).
\item Starting the loop again from step (ii).
\end{enumerate}

\begin{figure}
  \begin{center}
  \includegraphics*[width=\figsize]{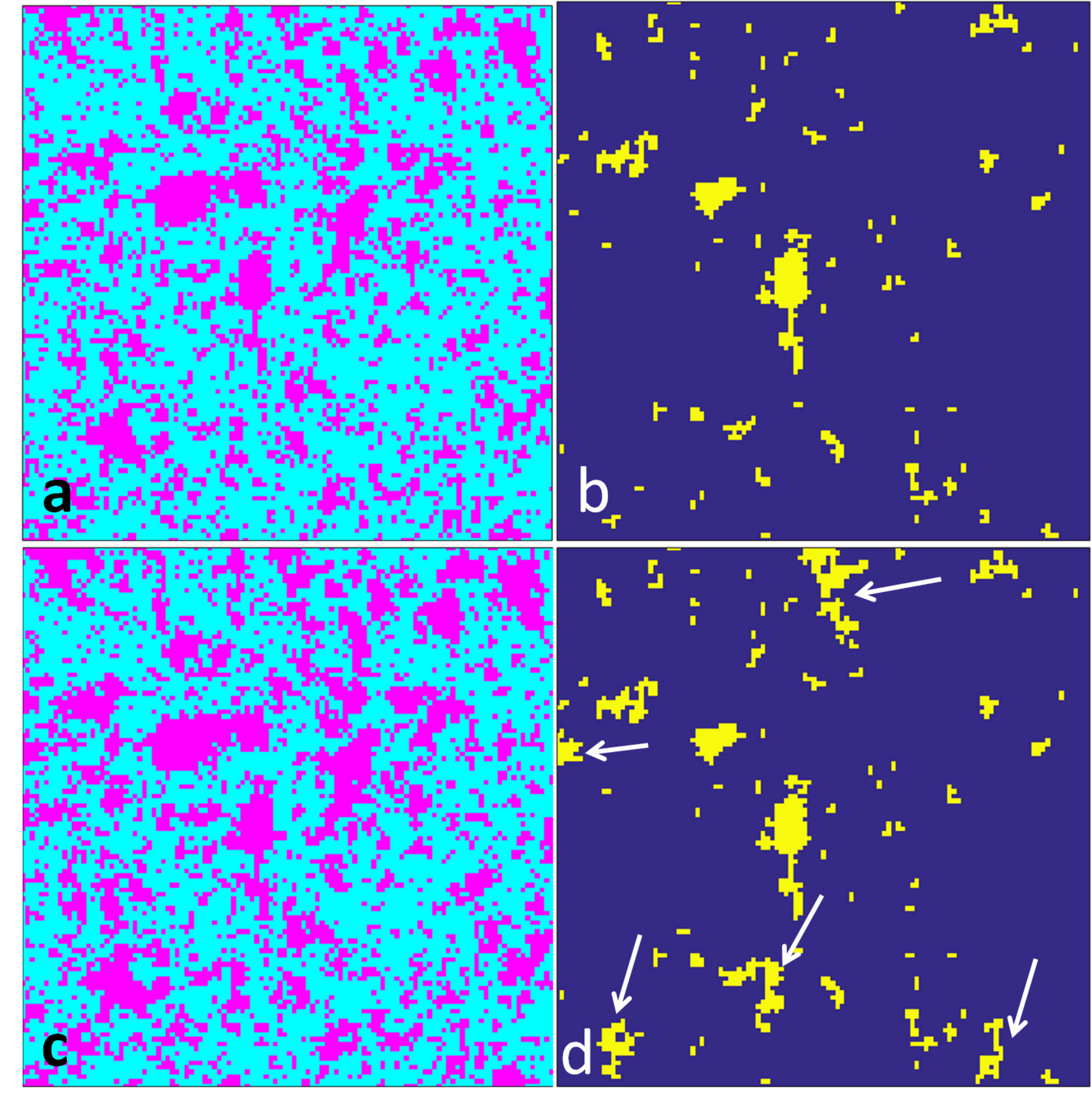}
  \end{center}
  \caption{Two successive iterations of the model's dynamics of the lower layer (panels a and c) and the corresponding upper layer (panels b and d). There are many ``holes'' (empty cells) in the lower layer but only some of them are expressed as sinkholes in the upper layer. Note the new large sinkholes (indicated by white arrows) that appeared in panel d compared to the former stage (panel b), and also note that the cavities in the lower layer continue to develop. Model's parameters: $\alpha=0.03$, $\gamma=0.06$, $p=0.01$.
 }
    \label{fig:lower}
\end{figure}

Note that according to these rules, the evolution process of both the upper and the lower layers is stochastic. Fig.~\ref{fig:lower} depicts an example of the two successive model iterations.
%illustrates the coupling between the two layers in the model.

\section{Results and Discussion}\label{sec:sec4}
\begin{figure}
  \begin{center}
  \includegraphics*[width=\figsize]{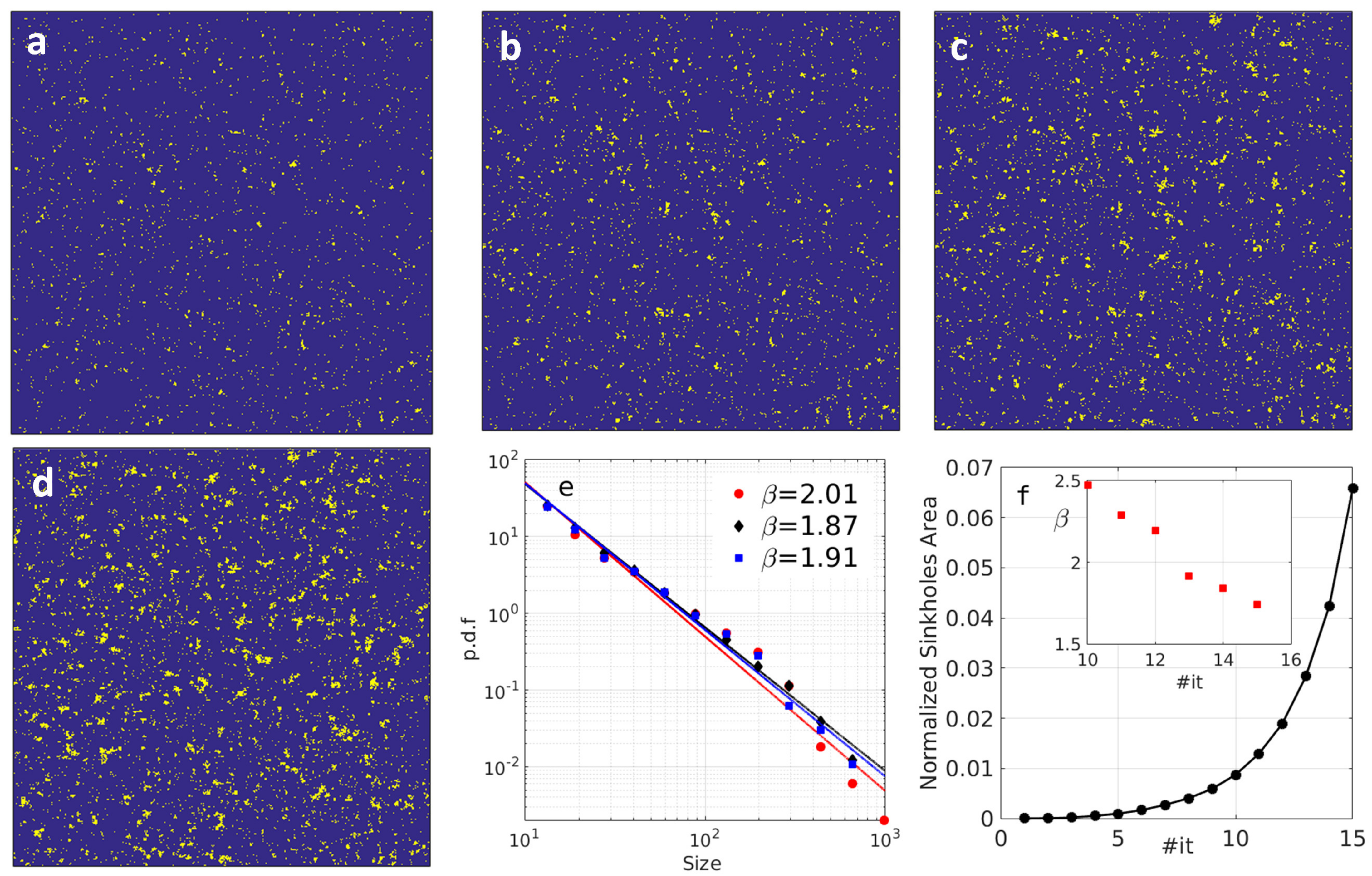}
  \end{center}
  \caption{
    (a-d) The evolution of the upper layer cavities (sinkholes) for iterations 12-15 in the entire $1024\times 1024$ grid---the area increases in existing sinkholes and new small sinkholes appear in each iteration. (e) Power-law distributions of sinkhole areas for iteration 15 of three realizations. The scaling exponent is close to two. (f) The normalized sinkhole area as a function of iteration and the calculated scaling exponent (inset) for the last six iterations. $\beta$ decreases since pdfs flatten in the more advanced iterations. Parameters: as in Fig.~\ref{fig:lower}.}
    \label{fig:fig3}
\end{figure}

The simulations were performed on a 1024 $\times$ 1024 grid, which is large enough for a good statistical analysis of the sinkholes.
The initial state in the lower layer is a random distribution of cavities with probability $p$; each initial configuration is considered as one realization.
Fig.~\ref{fig:fig3}a-d shows the evolution of the sinkholes (upper layer) for the four consecutive iterations. As expected, the total sinkhole area grows in each time step. The power-law distributions of three realizations are shown in Fig.~\ref{fig:fig3}e with scaling exponents between 1.87 and 2.01. Fig.~\ref{fig:fig3}f shows the increase in sinkhole area and the decrease of the scaling exponent $\beta$ for the last six iterations. It is interesting to note that $\beta$ decreases almost linearly from $\beta=2.47$ at iteration 11 to $\beta=1.74$ at iteration 15. This trend is due to the increase in the number of large sinkholes and the decrease in the number of small ones, a fact that ``flattens'' the pdf. We observe a decrease in $\beta$ with time for the Hever south and Tzuria sites (Fig. S8 ), although the error bars associated with this decrease are too large to assess the significance.

\begin{figure}
  \begin{center}
  \includegraphics*[width=4in]{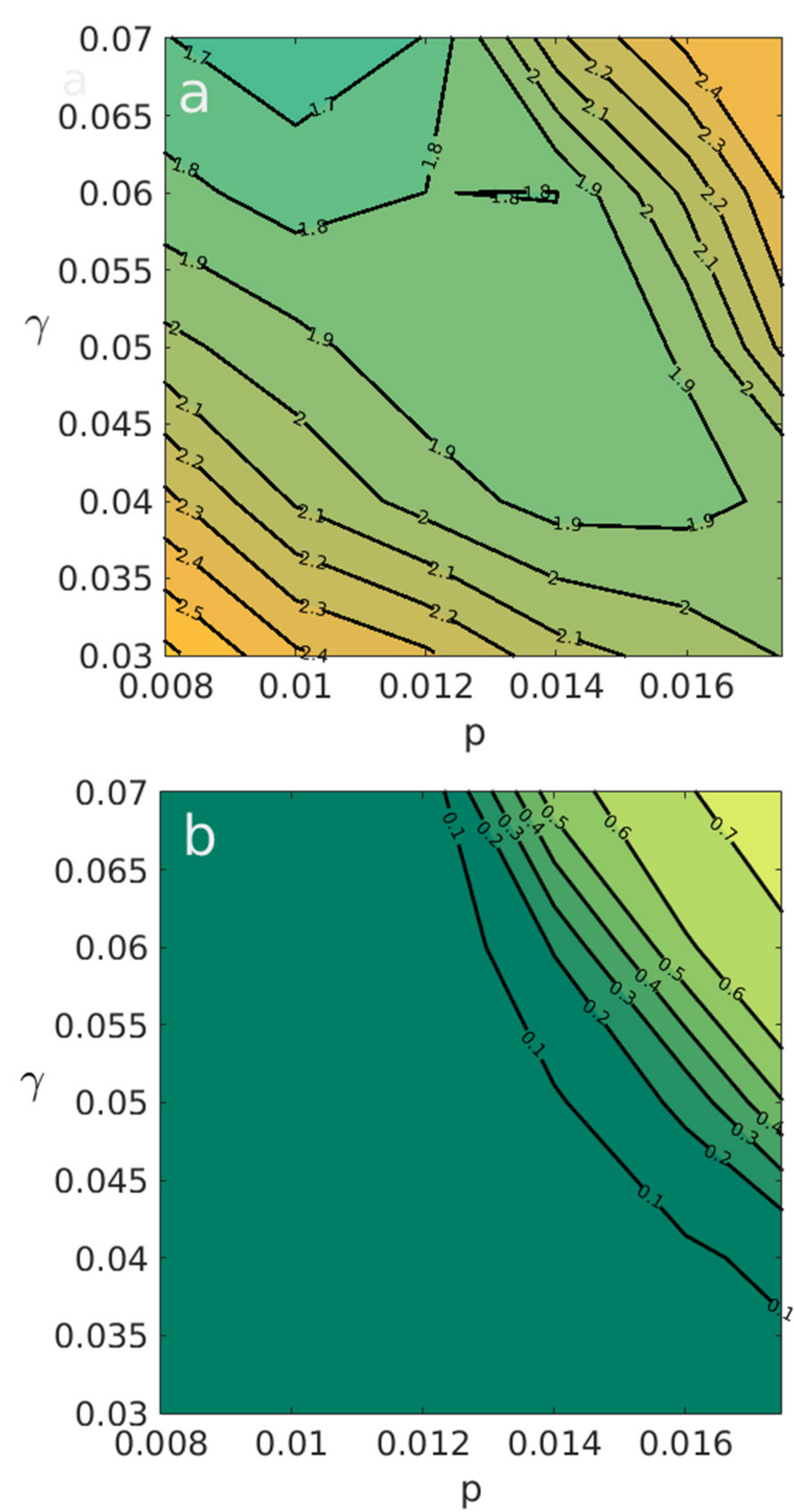}
  \end{center}
  \caption{ (a) The exponent $\beta$ as a function of $p$ and $\gamma$ for $\alpha=0.01$ and for iteration 15, for the regime in which the scaling exponent is around two. (b) The corresponding normalized sinkhole area.}
  \label{fig:fig4}
\end{figure}

Fig.~\ref{fig:fig4} depicts the scaling exponent, $\beta$, and the sinkhole normalized area as a function of $p$ and $\gamma$ for iteration 15 in the range where both $\beta$ and the sinkhole area approximately fit the observed data of the Dead Sea sinkholes (Fig.~\ref{fig:fig2}). $\beta$ decreases with $\gamma$ since when the probability for a site to collapse increases, larger sinkholes likely exist, causing the pdf to be flatter. In addition, the sinkhole area also increases in time, so the scaling exponent should also decrease in agreement with the field data (Fig. S8). Eq.~\ref{eq:cover} can be used to extrapolate the sinkhole area growth for the next few years; such an extrapolation is shown in Fig. S11.

The simple model we suggest above reproduces fairly well the observed scale-free distribution of sinkholes along the Dead Sea shoreline. The underlying physical mechanism for this scale-free behavior is the positive feedback between the cavity size and its growth modeled by Eq.~\ref{eq:dp} and by the collapse probability described by Eq.~\ref{eq:pA}. This positive feedback mechanism is similar to one of most known mechanisms for generating power-law distributions---the Yule process~\cite[][]{Newman-2005:power}. This effect that causes a big cavity to grow faster than a smaller cavity was dubbed the ``rich-get richer'' mechanism and can produce a distribution that follows a power law in its tail~\cite[][]{Simon-1955:class}. It can be shown analytically that the Yule process, which depends on three parameters, produces a power-law distribution with scaling exponent $2<\beta\leq 3$. In the model presented here, the dynamics is more complex than the standard Yule process, and the power-law distribution of cavities is expressed in the power-law distribution of aboveground sinkholes.

The continuing decline of the Dead Sea water level~\cite[][]{Yechieli-Abelson-Bein-Crouvi-Shtivelman-2006:sinkhole} will probably accelerate sinkhole formation since in addition to the dissolution of the salt layer by increased groundwater flow, the erosion process of poorly consolidated sediments interbedded within the salt deposits will increase the rate of cavity formation~\cite[][]{Kottmeier-Agnon-Al-Halbouni-2016:new}. Much remains for future studies to investigate, for example, whether the sinkhole size distribution on the Jordanian side of the Dead Sea also follows a power law. There are other sinkhole regions around the world, including the Barbastro--Balaguer salt anticline in northeastern Spain~\cite[][]{Lucha-et-al-2008:environmental} and the Sivas region in central Turkey~\cite[][]{Yilmaz-2007:GIS}. Although the number of sinkholes in these regions is much smaller than in the Dead Sea region, it is certainly possible that the pdfs of these sinkhole areas also follow scale-free behavior as on the Dead Sea shoreline.

%%% End of body of article:

%%%%%%%%%%%%%%%%%%%%%%%%%%%%%%%%
%% Optional Appendix goes here
%
% \appendix resets counters and redefines section heads
% but doesn't print anything.
% After typing  \appendix
%
% \section{Here Is Appendix Title}
% will show
% Appendix A: Here Is Appendix Title
%
%%%%%%%%%%%%%%%%%%%%%%%%%%%%%%%%%%%%%%%%%%%%%%%%%%%%%%%%%%%%%%%%
%
% Optional Glossary or Notation section, goes here
%
%%%%%%%%%%%%%%
% Glossary is only allowed in Reviews of Geophysics
% \section*{Glossary}
% \paragraph{Term}
% Term Definition here
%
%%%%%%%%%%%%%%
% Notation -- End each entry with a period.
% \begin{notation}
% Term & definition.\\
% Second term & second definition.\\
% \end{notation}
%%%%%%%%%%%%%%%%%%%%%%%%%%%%%%%%%%%%%%%%%%%%%%%%%%%%%%%%%%%%%%%%
%
%  ACKNOWLEDGMENTS

\begin{acknowledgments}
  We thank the Dead Sea Works and the Tamar Regional Council for their financial support and Meir Abelson, Gidi Baer, Noam Weisbord, and Yossi Yechieli for helpful discussions. We also thank two reviewers: Antonello Provenzale and an anonymous reviewer for their valuable comments. The data used are listed at:
  \begin{verbatim}
  http://www.boker.org.il/meida/negev/desert_biking/sinkholes/sinkholes-001.htm.
  \end{verbatim}
  The model's code is available upon request from the corresponding author.
\end{acknowledgments}

%% ------------------------------------------------------------------------ %%
%%  REFERENCE LIST AND TEXT CITATIONS
%
% Either type in your references using
% \begin{thebibliography}{}
% \bibitem{}
% Text
% \end{thebibliography}
%
% Or,
%
% If you use BiBTeX for your references, please use the agufull08.bst file (available at % ftp://ftp.agu.org/journals/latex/journals/Manuscript-Preparation/) to produce your .bbl
% file and copy the contents into your paper here.
%
% Follow these steps:
% 1. Run LaTeX on your LaTeX file.
%
% 2. Make sure the bibliography style appears as \bibliographystyle{agufull08}. Run BiBTeX on your LaTeX
% file.
%
% 3. Open the new .bbl file containing the reference list and
%   copy all the contents into your LaTeX file here.
%
% 4. Comment out the old \bibliographystyle and \bibliography commands.
%
% 5. Run LaTeX on your new file before submitting.
%
% AGU does not want a .bib or a .bbl file. Please copy in the contents of your .bbl file here.
\bibliographystyle{agufull04}
%\bibliography{all_sinkholes}

%Reference citation examples:

%...as shown by \textit{Kilby} [2008].
%...as shown by {\textit  {Lewin}} [1976], {\textit  {Carson}} [1986], {\textit  {Bartholdy and Billi}} [2002], and {\textit  {Rinaldi}} [2003].
%...has been shown [\textit{Kilby et al.}, 2008].
%...has been shown [{\textit  {Lewin}}, 1976; {\textit  {Carson}}, 1986; {\textit  {Bartholdy and Billi}}, 2002; {\textit  {Rinaldi}}, 2003].
%...has been shown [e.g., {\textit  {Lewin}}, 1976; {\textit  {Carson}}, 1986; {\textit  {Bartholdy and Billi}}, 2002; {\textit  {Rinaldi}}, 2003].

%...as shown by \citet{jskilby}.
%...as shown by \citet{lewin76}, \citet{carson86}, \citet{bartoldy02}, and \citet{rinaldi03}.
%...has been shown \citep{jskilbye}.
%...has been shown \citep{lewin76,carson86,bartoldy02,rinaldi03}.
%...has been shown \citep [e.g.,][]{lewin76,carson86,bartoldy02,rinaldi03}.
%
% Please use ONLY \citet and \citep for reference citations.
% DO NOT use other cite commands (e.g., \cite, \citeyear, \nocite, \citealp, etc.).

%% ------------------------------------------------------------------------ %%
%
%  END ARTICLE
%
%% ------------------------------------------------------------------------ %%
\end{article}
%
%
%% Enter Figures and Tables here:
%
% DO NOT USE \psfrag or \subfigure commands.
%
% Figure captions go below the figure.
% Table titles go above tables; all other caption information
%  should be placed in footnotes below the table.
%
%----------------
% EXAMPLE FIGURE
%
% \begin{figure}
% \noindent\includegraphics[width=20pc]{samplefigure.eps}
% \caption{Caption text here}
% \label{figure_label}
% \end{figure}
%
% ---------------
% EXAMPLE TABLE
%
%\begin{table}
%\caption{Time of the Transition Between Phase 1 and Phase 2\tablenotemark{a}}
%\centering
%\begin{tabular}{l c}
%\hline
% Run  & Time (min)  \\
%\hline
%  $l1$  & 260   \\
%  $l2$  & 300   \\
%  $l3$  & 340   \\
%  $h1$  & 270   \\
%  $h2$  & 250   \\
%  $h3$  & 380   \\
%  $r1$  & 370   \\
%  $r2$  & 390   \\
%\hline
%\end{tabular}
%\tablenotetext{a}{Footnote text here.}
%\end{table}

% See below for how to make sideways figures or tables.

\end{document}